\documentstyle[aps,preprint,floats,epsfig]{revtex}
\begin{document}
\def\bbox{{\,\lower0.9pt\vbox{\hrule \hbox{\vrule height 0.2 cm
\hskip 0.2 cm \vrule height 0.2 cm}\hrule}\,}}
\def\laq{\ \raise 0.4ex\hbox{$<$}\kern -0.8em\lower 0.62
ex\hbox{$\sim$}\ }
\def\gaq{\ \raise 0.4ex\hbox{$>$}\kern -0.7em\lower 0.62
ex\hbox{$\sim$}\ }


\def\a{\alpha}
\def\b{\beta}
\def\g{\gamma}
\def\G{\Gamma}
\def\d{\delta}
\def\D{\Delta}
\def\e{\epsilon}
\def\h{\hbar}
\def\ve{\varepsilon}
\def\z{\zeta}
\def\t{\theta}
\def\vt{\vartheta}
\def\r{\rho}
\def\vr{\varrho}
\def\k{\kappa}
\def\l{\lambda}
\def\L{\Lambda}
\def\m{\mu}
\def\n{\nu}
\def\o{\omega}
\def\O{\Omega}
\def\s{\sigma}
\def\vs{\varsigma}
\def\S{\Sigma}
\def\vphi{\varphi}
\def\av#1{\langle#1\rangle}
\def\pa{\partial}
\def\na{\nabla}
\def\hg{\hat g}
\def\un{\underline}
\def\ov{\overline}
\def\cF{{{\cal F}}}
\def\cG{{{\cal G}}}
\def\Hsl{H \hskip-8pt /}
\def\Fsl{F \hskip-6pt /}
\def\cFsl{\cF \hskip-5pt /}
\def\ksl{k \hskip-6pt /}
\def\pasl{\pa \hskip-6pt /}
\def\tr{{\rm tr}}
\def\tcF{{\tilde{{\cal F}_2}}}
\def\tg{{\tilde g}}
\def\shalf{\frac{1}{2}}
\def\nn{\nonumber \\}
\def\w{\wedge}
\def\ra{\rightarrow}
\def\la{\leftarrow}
\def\be{\begin{equation}}
\def\ee{\end{equation}}
\newcommand{\brr}{\begin{eqnarray}}
\newcommand{\bq}{\begin{equation}}
\newcommand{\err}{\end{eqnarray}}
\newcommand{\eq}{\end{equation}}
\newcommand{\nnu}{\nonumber}
\newcommand{\der}{\partial}


\def\cmp#1{{\it Comm. Math. Phys.} {\bf #1}}
\def\cqg#1{{\it Class. Quantum Grav.} {\bf #1}}
\def\pl#1{{\it Phys. Lett.} {\bf B#1}}
\def\prl#1{{\it Phys. Rev. Lett.} {\bf #1}}
\def\prd#1{{\it Phys. Rev.} {\bf D#1}}
\def\prr#1{{\it Phys. Rev.} {\bf #1}}
\def\prb#1{{\it Phys. Rev.} {\bf B#1}}
\def\np#1{{\it Nucl. Phys.} {\bf B#1}}
\def\ncim#1{{\it Nuovo Cimento} {\bf #1}}
\def\jmp#1{{\it J. Math. Phys.} {\bf #1}}
\def\aam#1{{\it Adv. Appl. Math.} {\bf #1}}
\def\mpl#1{{\it Mod. Phys. Lett.} {\bf A#1}}
\def\ijmp#1{{\it Int. J. Mod. Phys.} {\bf A#1}}
\def\prep#1{{\it Phys. Rep.} {\bf #1C}}


\title{String Universality}

\author{R. Brustein$^{(1)}$ and S. P. de Alwis$^{(2)}$}
\address{(1) Department of Physics,
Ben-Gurion University, Beer-Sheva 84105, Israel \\ (2) Department
of Physics, Box 390, University of Colorado, Boulder, CO 80309.\\
{\tt e-mail: (1) ramyb@bgumail.bgu.ac.il (2)
dealwis@pizero.colorado.edu}}
 \preprint{\vbox{COLO-HEP-442 \\ \
hep-th/0002087}} \date{January 2000}

\maketitle

\begin{abstract}

If there is a single underlying ``theory of everything"  which in
some limits of its ``moduli space" reduces to the five weakly
coupled string theories in 10D, and 11D SUGRA, then it is possible
that all six of them have some common domain of validity and that
they are in the same universality class, in the sense that the 4D
low energy physics of the different theories is the same. We call
this notion String Universality. This suggests that the true
vacuum of string theory is in a region of moduli space equally far
(in some sense) from all perturbative theories, most likely around the self-dual
point with respect to duality symmetries connecting them.  We
estimate stringy non-perturbative effects from wrapped brane
instantons in each perturbative theory, show how they are related
by dualities, and argue that they are likely to lead to moduli
stabilization only around the self-dual point. We argue that
moduli stabilization should occur near the string scale, and SUSY
breaking  should occur at a much lower intermediate scale, and that it
originates from different sources. We discuss the problems of moduli stabilization
and SUSY breaking in currently popular scenarios, explain why
these problems are generic, and discuss how our scenario can evade
them. We show that String Universality is not inconsistent with
phenomenology but that it is in conflict with some popular versions
of brane world scenarios.
\end{abstract}
\pacs{PACS numbers: 11.25Mj, 11.25-w, 04.65+e}

\renewcommand{\thefootnote}{\arabic{footnote}}
\setcounter{footnote}{0}

\section*{Contents}
\noindent
 1. Introduction\\
 2. Perturbative string theories\\
 3. String/M-theory  non-perturbative effects\\
 4. Moduli stabilization and SUSY breaking
\indent

 A. No-scale models

 B. Race track models

 C. Stabilization near the self-dual point\\
5. Consistency of string universality with phenomenology:
 \indent

 A. The value of $\rho$ in the Horava-Witten Theory

 B. Type I/IIB orientifold compactifications and Brane worlds

\section{Introduction}

It is well known that many fundamental questions in string theory,
such as the question of moduli stabilization, are beyond the reach
of perturbative techniques. Recent progress in elucidating the
duality relations between the different perturbative theories has
not resolved  these questions, but has enabled them to be posed in
a sharper fashion.

For instance,  if in a given perturbative string theory  the
string coupling constant is small $g<1$, so that perturbation
theory is valid, no potential  can be generated for the dilaton,
so that the coupling constant which is the vacuum expectation
value of the exponential of the dilaton is not fixed.
\footnote{See for example chapter 18 (pp 359-362) of \cite{jp} and
references therein.} S-duality on the other hand tells us that the
strong coupling region $g>1$,  of this string theory also has a
perturbative description in terms of  the S-dual theory, so, by
the same argument, no potential is generated in this theory. Thus
one  should expect that the string coupling is stabilized in a
region which is inaccessible to perturbative calculation from
either of the S-dual theories, in other words, it is fixed at an
intermediate value,  at or around the self-dual point $g\sim 1$.
\footnote{A similar idea was proposed in \cite{dns,drt} and by
Veneziano \cite{venezia}. Our arguments are similar in spirit but
different in detail. } Similar arguments can be made for the
moduli governing the sizes of the compact dimensions (T-moduli),
namely that they are fixed at  or around the self-dual point under
T-duality. This is because if the size of some compact dimensions
is much larger than the string scale then the theory is
effectively 5 or more dimensional, and no potential is generated
for the size moduli, since no potential for the moduli is allowed
in supergravity theories in more than four dimensions
\cite{bd2}\footnote{We ignore here the so-called gauged
supergravity theories  which yield AdS spaces as vacuum
configurations since a small 6-space would imply a large 4D
cosmological constant. They may however be relevant to brane world
scenarios.}. If the size of some compact dimensions is smaller
than the string scale then one takes the T-dual theory and makes
the same argument. Thus internal dimensions should  be fixed
around the self-dual point.

\begin{figure}[h]
\centerline{\psfig{figure=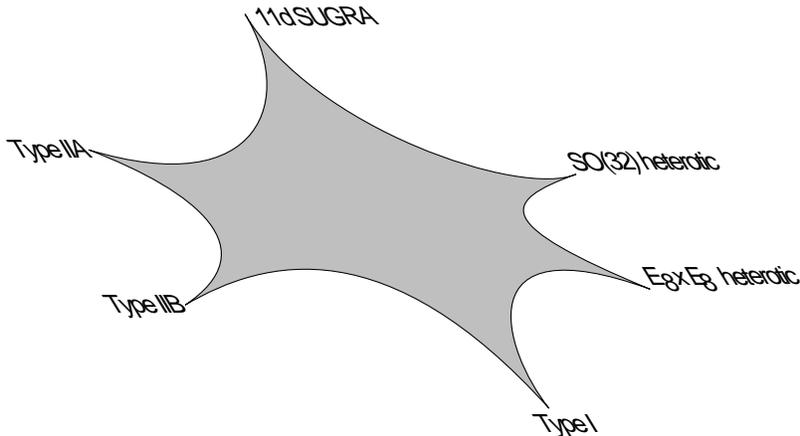,height=3.01in,width=4.2in}}
\caption{ {\label{fig1}} {``Star" diagram illustrating all string
theories, and 11D SUGRA as limits of one theory.} }
\end{figure}

Unfortunately, these self-dual points are in regions of minimal
computability,\footnote{See, for example, \cite{dine}.} they are
in some sense equally far away from all the perturbatively
computable string theories.  The region in which we expect the
true vacuum of string theory (with a fixed stable dilaton and
compact extra dimensions) to lie, is in the middle of the well
known star diagram (Figure \ref{fig1}) illustrating the unity of
the different perturbative string theories.

Our key assumption is that the four dimensional low energy theory,
which lives in the middle of the star diagram, should be in the
universal region of string/M theory. This means that all five
string theories and the 11D supergravity must be in the same
universality class, and the four dimensional low energy physics of
the different theories must be the same.  From this point of view,
the different perturbative string theories need not be any more
than different perturbative (physical) regularization schemes. In
particular, we assume that as one goes from the perturbative
region at any of the cusps to the center of the star diagram, the
infra-red spectrum that is common to all the different starting
points is unchanged.  For example, there should be a graviton in
this region since it exists in all the perturbative string
theories.\footnote{This is similar to the assumption made in
\cite{bd} in connection with the strongly coupled heterotic
string, but \cite{bd}   was written before the importance of
duality was fully realized. What we are proposing here amounts in
part to an attempt to extend the arguments of \cite{bd}  to all
perturbative string theories taking the dualities into
consideration.} As for the surviving gauge group, this is a more
complicated issue since different starting points have different
gauge groups, with recent developments, for instance F-theory,
giving a wide variety of groups.\footnote{It may be the case that
the gauge group that survives in the four dimensional theory in
the central region, is the common subgroup of all the starting
points. However, we would leave this for the moment as an open
question.}

Traditionally the view that has been taken is that the real world
is described by a single perturbative string model. In other
words, that for some unknown reason nature picks one weakly
coupled model over the others. Thus for instance up until recently
it was thought that the heterotic $E_8\times E_8$ (HE) theory was
the theory that describes the real world. The realization that the
different theories are just perturbative descriptions about
different points in moduli space has changed that perspective.
Nevertheless the traditional belief still survives in a modified
form;  for example, currently it has become fashionable to use the
phenomenology of type I theories (with D-branes) on the grounds
that they may be better descriptions of nature than the heterotic
string.

One reason for the popularity of type I models is that in weak
coupling heterotic theories one gets too small a value for the 4D
Planck mass $M_P$, when ``experimental" values for $\a_{YM}$ at
unification are used and additionally the compactification scale
$M_C$, is identified with the grand unification scale $M_{GUT}$.
The argument was made by Witten \cite{ew} in the heterotic SO(32)
(HO) - type I S-duality context, that one might replace weakly
coupled HO string theory $g_{HO}<1$, by weakly coupled type I
theory $g_I<1$ , in which case one could avoid this problem.
However our string universality conjecture would give a different
interpretation. The physics of the strongly coupled HO theory must
be equivalent to the physics of the weakly coupled type I theory,
where of course the dilaton cannot be stabilized, so that any
comparison to phenomenology is not meaningful. The true vacuum
should be around the self-dual coupling $g_{HO}\sim g_{I}\sim 1$.

In the HE case according to Horava and Witten\cite{hw}, the strong
coupling theory is (in the low energy limit) 11-dimensional
supergravity on $S_1/Z_2$ (HW).  The naive relations here would
seem to give the size of the interval in the eleventh direction
$\rho$ to be about 70 times the size of the six volume (and the
11D Planck scale). This has given rise to a picture where the
gauge couplings evolve according to the four dimensional gauge
theory picture but the gravitational coupling becomes five
dimensional at a point well below the unification point\cite{bd2}.
There is no possibility of such a picture arising in the HO/I
theories where as we have argued above, the discrepancy of scales
needs to be resolved with a coupling of order unity and there is
no room for a five dimensional scenario. According to our
conjecture of string universality the HE theory should give the
same low energy physics. So the naive M theory picture needs to be
revised. Our (preliminary) investigations show that with currently
available calculations of threshold corrections $g\sim O(1)$ is
indeed a viable scenario.

It seems unlikely that the strong coupling picture, which implies
that the 11D action is a good starting point for understanding 4D
physics, is a description of the real world, since within this
picture it is not possible to generate a potential which
stabilizes the moduli. Indeed the phenomenology of HW theory is
just a reparametrization of that coming from HE theory. In
particular, the potentials that have been obtained upon
compactification demonstrate the same runaway behavior as in the
weak coupling analysis \cite{nilles,ovrut}.

The calculation of the length of the eleventh dimension needs to
revised by allowing for an order unity numerical factor in the
relation between the Kaluza-Klein scale and the unification scale,
and also to allow for the analog of threshold corrections.  Indeed
the latter may be taken into account by using a result in Witten's
original paper \cite{ew}\footnote{These are the corrections that
one would identify as threshold corrections (in the context of the
field theory they are identified with certain Green-Schwarz
anomaly cancelation terms) in the 10D theory as pointed out in
\cite{bd2} and discussed in detail in \cite{sn}.}. If these
numerical factors are included and we use a nonstandard embedding
then a picture emerges where $\rho$ is of the order of the eleven dimensional Planck scale and the
compactification scale. In other words the distance between the boundaries is of the order of the quantum fluctuations  and no fifth dimension appears. This would then be
compatible with our universality hypothesis since in the HO/I
picture there is no room for a five dimensional picture below the
string scale. In all this one is assuming that the
compactification/unification scale is somewhat below, though
close, to the string scale so that the field theory approximation
makes sense.

Recently there has been much discussion within the context of
so-called brane-world scenarios that the string scale may be as
low as 1 TeV \cite{jl,AASD}.  While there is as yet no convincing
string model that accounts for all the requirements that need to
be met to have a viable description of this type, the string
models that might be considered as candidates for this are T-duals
of the type I theory. For example consider compactifying on a six
torus (or orbifold) and T-dualizing, in which case we get a IIB
orientifold with $2^6$ orientifold planes and 32 D3 branes on
which the standard model may live. Gravity however propagates in
the bulk as well.  If the string scale $l_s^{-1}$ is taken to be
around a TeV then in order to get the right value for the Planck
mass one needs to have the dimensions transverse to the D3 brane
to be very large compared to the string scale (about $10^3l_s$).
Now by our universality hypothesis the same low energy physics
must be seen from the U dual (S duality times T duality on a six
torus) HO theory. But in the latter gravity and gauge theory
propagate on the same space. The low energy theories are
compatible only if in the former (i.e. the brane world) the
transverse directions are of the order of the string scale. This
is of course consistent with the argument above that the
compactification scale should be around the string scale. However
this would mean that the string scale must be close to the 4D
Planck scale. Thus it seems that only the conventional view of the
size of the string scale is compatible with our hypothesis.  Of
course, it is  possible that these theories have to be considered
in terms of non-standard compactifications (leading to gauged
supergravity), but since it is not  clear to us whether such a
scenario has a viable string description we will not pursue this
question further in this paper.

Any of the five string theories and the Horava-Witten theory must
lead at low energies and in 4D to the same potential for the
moduli, assuming a compactification which results in N=1 SUSY in
4d. The low energy theory should have a SUSY breaking minimum with
vanishing cosmological constant, and 4D dilaton stabilized at such
a value that the (unified) gauge coupling is weak. On the other
hand, the 10D theory must have intermediate coupling so that
string perturbation theory has the opportunity to break down,
otherwise there is no way that a potential could have been
generated. We assume that at intermediate, or strong coupling, the
low energy 10D actions are in fact just determined by general
covariance supersymmetry and gauge invariance, and the actions for
the different string theories are obtained by field redefinitions.
Whatever mechanism gives rise to the 4D potential, our string
universality assumption is that it ought to be independent of the
particular perturbative string starting point.

General arguments based on PQ symmetries, and how they break due
to non-perturbative effects, show that the superpotential must be
a sum of exponentials in the moduli. These exponentials could be
generated by stringy or field theoretic non-perturbative effects.
For instance, the ``race-track" mechanism  for the stabilization
of moduli\footnote{See \cite{dine} for a recent review.} envisages
at least three exponential terms, coming, for instance from
gaugino-condensates \cite{gaugino1,gaugino2}, which can balance
against each other when all of them are small
\cite{krasnikov,multigaugino}. A constant term in the
superpotential while allowed by the symmetries has no natural
mechanism for its generation. The one known exception is when the
field strength of the antisymmetric two form field acquires a
vacuum expectation value but this is quantized in units of the
string scale \cite{rohmw} and hence yields too large a value for
the scale of the gauge theory.

In the absence of a constant in the superpotential, if we use the
Kahler potential of string perturbation theory, the moduli
potential will give only a weak local minimum with zero
cosmological constant while the global minimum has negative
cosmological constant \cite{bs}. One might think that this
conclusion is avoided in the no-scale type models \cite{noscale},
but these are valid only at tree level and are destabilized, for
example, by threshold corrections.  So unless one finds a
mechanism for generating a constant in the superpotential the only
way out of this is to assume that the Kahler potential is
drastically modified from its form in string perturbation theory
\cite{bd,npkahler}. One can now speculate as to how such
non-perturbative terms might arise (from wrapping of branes on
cycles in CY spaces for instance). Combined with constraints
coming from duality we find that it is extremely hard to generate
significant contributions. We propose therefore that the question
of moduli stabilization should be decoupled from SUSY breaking
(See also \cite{dns}).

In the next section  we review some known facts perturbative
string theories and their dualities, and set up notation. In
section 3 the origin of string non-perturbative effects and their
dualities are discussed. Section 4 is devoted to highlighting the
problems associated with currently popular scenarios for moduli
stabilization and SUSY breaking. In section 4  we propose as an
alternative a decoupling of the two issues. We suggest that moduli
are all stabilized at or near the self-dual point by string
non-perturbative effects while SUSY breaking happens at a much
lower scale perhaps as a result of field theoretic
non-perturbative effects.

\section{Perturbative string theories}

For clarity we review some well known issues first. The
perturbation expansion of any string effective action is given by
 \begin{equation}
\Gamma[\phi,
G_{\mu\nu},B_{\mu\nu},...]=\sum_ie^{-\phi_0\chi_i}S_i[\tilde\phi,
G_{\mu\nu},B_{\mu\nu},...]
 \ee
where the sum is over Riemann surfaces and $\chi_i=2-2h_i-b_i$ is
the Euler character of the surface with $h_i$ handles and $b_i$
boundaries and we have split $\phi =\phi_0+\tilde\phi$ where the
first term is the constant part of the dilaton defined so that
$\int\tilde\phi =0$.  This form of the effective action clearly
restricts the form of the potential for $\phi$. If one translates
$\phi\rightarrow\phi +\shalf\ln t$ then each term in the expansion
acquires a factor $t^{1-h_i-\shalf b_i}$ and the only potential
that would be allowed is of the form
 \begin{equation}
V(\phi )=\sum_i\L_ie^{-\phi\chi_i}.
 \ee
In superstrings with unbroken SUSY (formulated on a flat
background) $\L_i=0$ for all $i$\footnote{Rigorously proven up to
$i=2$ but expected to be true for all $i$.} which is of course a
necessary consistency condition.  But in general such a potential
is present in non-supersymmetric string theories (except that
$\L_{S_2}$ is zero) or superstrings with broken SUSY (say by the
Scherk-Schwarz mechanism). Assuming it exists, the critical point
$\phi =\phi_0$ (which should be such that $V(\phi_0) =0$) is
generically at $g=e^{\phi_0}\sim 1$ since the ratios of
coefficients of the perturbation series should be of order one. In
general the potential may be written as\footnote{$V$ will of
course depend on other moduli as well but we will ignore this for
the time being.}
 \begin{equation}
 V[\phi ]=\sum_i\L_ie^{-\phi\chi_i}+V_{np}[\phi],
 \ee
The non-perturbative term $V_{np}$ is expected to depend on
the coupling as $e^{-1/g}$ or $e^{-1/g^2}$. We will discuss later
the contributions to $V_{np}$ coming from brane-instanton effects.

Let us first list the low energy actions of the different string
theories  and their relations with each other. We only include the
dilaton-gravitational and the gauge couplings since our discussion
is going to be confined to the relations between these couplings.

The low energy effective action of type I string theory is the
following,
 \begin{eqnarray}
\label{typeIea}
 \Gamma_I&=&{1\over (2\pi)^7 l_I^8}
 \int_{M_{10}}[e^{-2\phi_I}\sqrt{-G_I}(R+4(\nabla\phi)^2)_I]\nn
 & &-{1\over 4(2\pi)^7l_I^6}\int_{M_{10}}e^{-\phi_I}
 \sqrt{ -G_I}\tr F^2_I.
 \end{eqnarray}
The low energy effective action of heterotic SO(32) string theory
(HO) is the following,
 \begin{eqnarray}
\label{HOea} \Gamma_{HO}&=& {1\over (2\pi)^7l_{HO}^8}
\int_{M_{10}} \sqrt{-G_{HO}}e^{-2\phi_{HO}}
 \left\{ R+4(\nabla\phi)^2 \right\}_{HO}\nn
 & &-{1\over 4(2\pi)^7l_{HO}^6}
 \int_{M_{10}}\sqrt{ -G_{HO}}e^{-2\phi_{HO}}\tr F^2_{HO}.
 \end{eqnarray}
The fields and parameters of these two theories which are S-dual
to each other are related by
\begin{eqnarray}
\label{HOsd}
 \phi_I&=&-\phi_{HO} \nonumber \\
 G_{\mu\nu,I}&=&g_{HO} e^{-\phi_{HO}}G_{\mu\nu,HO} \\
 g_{I}&=& e^{<\phi_{I}>_0}=e^{-<\phi_{H}>_0}=\frac{1}{g_H} \nonumber \\
 l_I^2&=&g_{HO}l_{HO}^2. \nonumber
\end{eqnarray}

The low energy effective action of heterotic $E8\times E8$ string
theory (HE) is the following,
\begin{eqnarray}\label{HEea}
\Gamma_{HE}=& &{1\over (2\pi )^7 l_{HE}^8} \int_{M_{10}}
\sqrt{-G_{HE}}e^{-2\phi_{HE}}
 \left\{R+4(\nabla\phi_{HE})^2 \right\}_{HE}
 \nonumber \\
  &-& \sum_i{1\over 4(2\pi)^7l_{HE}^6}\int_{M_{10}}
  \sqrt{-G_{HE}}e^{-2\phi_{HE}}\tr F_i^2.
\end{eqnarray}
The HO and HE theories are related by T-duality. So
$l_{HE}=l_{HO}$ and $G_{HE}=G_{HO}$ and if HO is compactified on a
circle of radius $R$ then the physically equivalent HE is
compactified on a circle of radius $l_{HE}^2/R$, with
$e^{\phi_{HE}}={l_{HO}\over R} e^{\phi_{HO}}$.

In the strong coupling limit the HE theory goes over to the
Horava-Witten (HW) \cite{hw} theory \footnote{For discussions of
the phenomenology of the Horava-Witten  theory see the  reviews
\cite{nilles,ovrut,munoz}.}, whose action is given by
 \begin{equation}
 \label{HWea}
 \Gamma_{HW}={1\over 2 \k_{11}^2}\int_{{\cal M}_{11}}
 d^{11}x\sqrt{G} R-{1\over8\pi (4\pi k_{11}^2)^{2/3}}
\left[\int_{{\cal M}_{10}} d^{10}x\sqrt{G}\tr F^2_1
 +\int_{{\cal M}_{10}} d^{10}x\sqrt{G}\tr F^2_2\right].
 \ee
The gauge fields in HW are the same as in the HE theory and the
metric is related by
 \begin{equation}\label{HEHWmet}
 ds^2_{HW}=e^{-{2\over 3} \phi_{HE}}G_{\mu\nu,HE} dx^{\mu}dx^{\nu}
 + e^{{4\over 3}\phi_{HE}}dy^2,
 \ee
where $y$ is the eleventh coordinate. The parameters of the two
theories are related by
 \be
 \label{HWHE}l_{11}=l_{HE}g_{HE}^{1/3},~~\rho =l_{HE}g_{HE}
 \ee
where we put $2\k_{11}^2=(2\pi)^8l_{11}^9$ and $\int dx^{11}\sqrt
G_{11}=2\pi\rho$.

Let us now make the following reasonable assumptions about the low
energy effective actions of all string theories and 11D SUGRA.
These assumptions are usually made by most authors in superstring
theory though they are not always explicitly stated.
\begin{itemize}
\item
There exists a minimum of the effective action that breaks
supersymmetry with zero cosmological constant.
\item
The low energy effective action can be written in terms of the
perturbative spectrum of low energy fields even in the regime
where perturbation theory is formally invalid, except that some
fields (moduli) which are not protected by gauge symmetries will
acquire a potential and become massive. In particular, there will
be massless graviton and gauge fields that will couple exactly as
expected from the perturbative calculations because of general
covariance and gauge invariance.
\end{itemize}

The first assumption is certainly non-controversial, but can be
posed in two different degrees of severity. The weaker variant
that we call the ``practical cosmological constant problem", is
the  one of ensuring that to a given accuracy
within a given model the cosmological constant vanishes. Stated
differently, it is the requirement that models should allow a
large universe to exist with reasonable probability, and have 4D
flat space as a solution. We believe that this requirement should
be imposed on any model. The stronger variant is the general
question of why  the cosmological constant today is so
small in natural units \cite{weinberg}. This question is
especially interesting in light of the hints that experimentally a
small non-vanishing cosmological constant seems to be favored. Of
course, the resolution of this issue is extremely important, but
is outside the scope of many models, and we believe that it should
not be absolutely required from models.

As for the second assumption, one might think that the spectrum of
the theory at strong coupling is completely different from the
spectrum at weak coupling. This is the case, for instance, in QCD,
though even there the fundamental theory is still written in terms
of the quarks and gluons\footnote{The latter point appears to call
into question the necessity of formulating M theory in terms of
fundamental degrees of freedom that are different from the
perturbative states of string theory. In the view expressed here
all the non-perturbative objects - branes, Black holes etc. - are
the analogs of instantons and solitons in field theory.}. This is
a possibility that cannot be ruled out at this stage of
development of string theory, but existing indications (that we
outline below) suggest that it is at least plausible that there
are no phase transitions on the way from weak to strong coupling.

Perturbation theory establishes the form of the low energy action
in the two extreme regions which are related by field
redefinitions. For instance if $\phi_{HO}$ is the HO dilaton then in
the region $\phi_{HO}\ll -1$ we would have a perturbative heterotic
description of the theory with perturbative gravity and gauge
dynamics, and  a type I description in the region $\phi_{HO}\gg +1$,
with perturbative gravity and gauge dynamics.  The region of coupling
in which we cannot be sure that perturbation theory of at least one of the
theories I and HO is valid is probably quite small (say, $0.7<g_{HO}<1.5$).
In going from one side of this region to the other we get low energy
4D theories that are just field redefinitions of each other.
Thus one might expect that at intermediate values of the coupling
the nature of the theory is not radically altered. A similar argument
holds for the HE/HW duality. Although HW theory has no dilaton
(and there is no perturbative analog as in type I)  the physical
distance $\rho$ along the transverse eleventh direction plays the
role of the dilaton. Whatever the underlying theory is, in the
region where all length scales including $\rho$ are  greater than
$l_{11}$ 11D SUGRA should be valid.
This is a strong coupling version of the heterotic string
effective action but there is no trace of any non-perturbative
effects or phase transitions as is evident from the fact that the
four dimensional effective action coming from HW is basically the
same as that coming from string theory with appropriate field
redefinitions. Thus again one might argue that the nature of the intermediate region is not very different from what these boundary regions would suggest except of course that the moduli would
have a potential. One would then hope to extract some information
about the central region from the physics of the boundary region.

\section{String/M theory Non-perturbative effects}

String universality suggests that not only are different perturbative
effective actions  related, but also non-perturbative induced
terms in these actions are universal. After all, much of the low
energy physics of any string theory is determined by
non-perturbative interactions. A natural source of such universal
non-perturbative effects are BPS brane-instantons.

Motivated by string universality we propose that string/M
theoretic non-perturbative effects (SNP) originate from
supersymmetric BPS branes \cite{Witten1:1995}.\footnote{Recently
Sen \cite{sen} has discussed non BPS branes, which deserve a
separate discussion. Recently it has been argued \cite{harvey} that these correspond to sphelarons in field theory. But
we do not expect our qualitative
conclusions on moduli stabilization to be significantly modified by these effects.} The branes that we
consider are objects which are point like in Euclidean four space
and are obtained by wrapping the extended directions of the brane
(including its Euclidean time direction) around some cycle in the
compact space. Their action is the product of brane tension and
the volume of the cycle around which its world volume is wrapped.
Since under dualities, BPS branes transform into BPS branes, SNP
in one string theory (or 11d SUGRA) transform into SNP in the
other theory. The matching between BPS states in theories dual to
each other is complete, and is actually considered one of the
central pieces of evidence for the ``correctness" of dualities.
Therefore, we argue that we can make a complete list of SNP, by
taking into account the known BPS branes, and that there's no room
for additional SNP which are sometimes assumed in some models.
Such complete matching is obviously compatible with
non-perturbative string universality.

We focus for simplicity on two moduli, the volume of the 6 compact
dimensions $V$, and the string coupling $g$ (or the size of the
11d interval $\rho$ in Horava-Witten theory). In accordance with
the general arguments about the dependence of SNP  on moduli, we
find that SNP depend exponentially on $V$ and on $1/g$ (or
$1/g^2$). In the weak coupling, large volume region of moduli
space, which we call the ``boundary region" of moduli space,
dominant SNP come from two or sometimes three BPS branes. The
dominant SNP come from the brane-instantons whose actions have the
weakest dependence on the string coupling $g$, and
compactification radius $R$, or their product. Therefore for
drawing conclusions about the boundary region of moduli space it
is enough to consider only very few contributions.

For each theory we compare the 4d Yang-Mills coupling
$\alpha_{YM}= \frac{g_{YM}^2}{4 \pi}$, and Newton's constant
$G_N$, and express the two couplings in terms of the string
coupling and string scale of each string theory. We formally
define the volume of the 6D compact manifold,  $V=R^6$, and assume
for simplicity that Euclidean time extension is also given by $R$.
Obviously more complicated possibilities can be considered, but we
leave them to future work. In this section we will omit most
numerical factors, $\pi$'s etc., since they will not be important
for our purposes. The necessary numerical factors have appeared in
previous sections or can be computed in a straightforward manner.
SNP  are simply given $e^{- {\rm action}}$ (ignoring the prefactor
and questions related to fermion zero modes). In the following the
branes are denoted by standard notation, $F$/$D$ stands for a
fundamental/Dirichlet brane\footnote{For uniformity of notation we
have denoted all branes which are conventionally called NS,
because they are sources for NS-NS fields, by the letter F
(conventionally used only for the NS 1-brane). After all D branes
are not called ``R branes".}, followed by the number of spatial
dimensions of the brane.

\subsection{Type I vs. heterotic $SO(32)$}
In type I  and HO, the 4d Yang-Mills coupling  $\alpha_{YM}=
\frac{g_{YM}^2}{4 \pi}$, and Newton's constant $G_N$ are given by
table \ref{tbcIHO}.
\begin{table}[h]
    \begin{center}
\begin{tabular}{||c||c|c||}
    \hline \hline
     &     Type I &  HO     \\
  [.3ex]  \hline\hline
      $ \alpha_{YM} $ & $ g_I \left(\frac{l_I}{R}\right)^6$
     & $g_{HO}^2 \left(\frac{l_{HO}}{R}\right)^6$  \\
   [.3ex]  \hline
    $ G_N $ & $ g_I^2 \left(\frac{l_I}{R}\right)^6 l_I^2$
     & $g_{HO}^2 \left(\frac{l_{HO}}{R}\right)^6 l_{HO}^2$  \\
         [.3ex] \hline\hline
         \end{tabular}
   \caption{$\alpha_{YM}$ and $ G_N $ in type I/HO.}
  \label{tbcIHO}
\end{center}
\end{table}
The $S$-duality relations between the two I-HO couplings are given
in equations (\ref{HOsd}). In table \ref{tbIHO} we list the type
of branes and their actions in the two $S$-dual theories, type I,
and HO. In each row the branes are related by duality, and their
actions are related by (\ref{HOsd}).
\begin{table}[h]
    \begin{center}
    \begin{tabular}{||c|c|c||c|c||}
    \hline \hline
     & \ \ \ Type I \ \ \  &   & \ \ \ HO  \ \ \  &   \\
    \hline\hline
     & $ brane $ &\ \ \  action \ \ \ & brane & \ \ \ action \ \ \ \\
   [.5ex]  \hline
     & $ D5$ & $
    \frac{1}{g_I} \left(\frac{R}{l_{I}}\right)^6  $
    & $ F5 $ &  \
    $\frac{1}{g_{HO}^2} \left(\frac{R}{l_{HO}}\right)^6 $ \  \\
    [.5ex] \hline \hline
     & $ D1 $  &
    $ \frac{1}{g_I} \left(\frac{R}{l_{I}}\right)^2  $
    & $ F1$  & $ \left(\frac{R}{l_{HO}}\right)^2 $ \\
       [.5ex] \hline\hline
  \end{tabular}
  \caption{BPS branes and actions in type I/HO.}
  \label{tbIHO}
\end{center}
  \end{table}
In type I there are only D-branes, and in HO only the heterotic
string and 5-brane.

\subsection{ $M$ on $R^{10} \times S^1/Z_2$ vs. heterotic $E8\times E8$}

In HE  and HW theories, the 4d Yang-Mills coupling  $\alpha_{YM}=
\frac{g_{YM}^2}{4 \pi}$, and Newton's constant $G_N$ are given by
table \ref{tbcHEHW}.
\begin{table}[h]
    \begin{center}
\begin{tabular}{||c||c|c||}
    \hline \hline
     &     HE &  HW     \\
  [.5ex]  \hline\hline
    $\ \alpha_{YM}\ $ & $ g_{HE}^2 \left(\frac{l_{HE}}{R}\right)^6$
     &  $\left(\frac{l_{11}}{R}\right)^6$ \\
   [.5ex]  \hline
    $ G_N $ & $\  g_{HE}^2 \left(\frac{l_{HE}}{R}\right)^6
  l_{HE}^2 \ $ & $\ \frac{1}{\rho} \left(\frac{l_{11}}{R}\right)^6
  l_{11}^3 \  $ \\
         [.5ex] \hline\hline
\end{tabular}
   \caption{$\alpha_{YM}$ and $ G_N $ in HE/HW.}
  \label{tbcHEHW}
\end{center}
\end{table}
The duality relations between the two HE-HW couplings are
 \begin{eqnarray} \label{HEHWduality}
 l_{HE}\ g_{HE} &=& \rho  \nonumber \\
 l_{HE} g_{HE}^{1/3} &=& l_{11}  \\
 l_{HE}^2 &=& \frac{ l_{11}^3}{\rho}, \nonumber
 \end{eqnarray}
where only two are independent.

In  table \ref{tbHEHW} we list the type of branes and their
actions in the two theories, HE, and HW. In each row the branes
are related by duality, and their actions are related by
(\ref{HEHWduality}). In HE there are only the F string and its
magnetic dual the F 5-brane. $M$ stands for an $M$-theory brane,
followed by the number of spatial dimensions of the brane. We need
to discuss two possible orientations of each type of brane, one
where one direction is longitudinal to the 11th dimension and the
other where all brane directions are transverse to it. Only one
possible orientation of each $M$-brane has a dual and therefore
(we conjecture) is allowed. The other two possibilities, $M5$
longitudinal and $M2$ transverse\footnote{These would correspond
to $D_4$ and $D_2$ branes in the string theory limit and are
present in IIA but are absent in the heterotic theories.},  are
(we believe) absent (i.e. their pre-factors should vanish) since
in the S-dual HE theory they are absent. In  the HW theory this
happens at the boundaries because the fields  that they couple to
are odd under the $Z_2$, $x_{11}\ra -x_{11}$ and are projected out
there.\begin{table}[h]
    \begin{center}
    \begin{tabular}{||c|c|c||c|c||}
    \hline \hline
     & \ \ \ HE \ \ \  &   & \ \ \ HW  \ \ \  &   \\
    \hline\hline
     & $ brane $ &\ \ \  action \ \ \ & brane & \ \ \ action \ \ \ \\
   [.6ex]  \hline
 \  \ & $ F5 $ & $ \frac{1}{g_{HE}^2} \left(\frac{R}{l_{HE}}\right)^6$
    &\ $ M5 $ transverse \ & $ \left(\frac{R}{l_{11}}\right)^6$ \  \\
    [.6ex] \hline \hline
   \  \ &\ $ F1 $\ & $ \left(\frac{R}{l_{HE}}\right)^2  $
    &\ $ M2 $ longitudinal\  & $
    \left(\frac{R}{l_{11}}\right)^2 \frac{\rho}{l_{11}} $ \\
       [.6ex] \hline\hline
  \end{tabular}
  \caption{BPS branes and actions in HE/HW.}
  \label{tbHEHW}
\end{center}
  \end{table}

\subsection{ $M$ on $R^{10} \times S^1$ vs. type IIA}

In $M$ on $R^{10} \times S^1$ (MS1)  and type IIA theories, the
comparison of the 4d Yang-Mills coupling is meaningless since
there are no perturbative gauge fields in these theories. Newton's
constant $G_N$ is given by table \ref{tbcIIAMS1}.
\begin{table}[h]
    \begin{center}
\begin{tabular}{||c||c|c||}
    \hline \hline
     &     IIA &  MS1     \\
  [.5ex]  \hline\hline
 $ G_N $ & $\  g_{IIA}^2 \left(\frac{l_{IIA}}{R}\right)^6
 l_{IIA}^2  \ $ & $\ \frac{1}{\rho} \left(\frac{l_{11}}{R}\right)^6
  l_{11}^3 \  $ \\
         [.5ex] \hline\hline
\end{tabular}
   \caption{$ G_N $ in IIA/M.}
 \label{tbcIIAMS1}
\end{center}
\end{table}
 The duality relations between the two IIA, HW couplings are
\begin{eqnarray} \label{IIAMS1duality}
l_{IIA}\ g_{IIA} &=& \rho \nonumber \\
 l_{IIA} g_{IIA}^{1/3} &=& l_{11} \\
 l_{IIA}^2 &=& \frac{ l_{11}^3}{\rho}, \nonumber
\end{eqnarray}
where only two are independent.

In  table \ref{tbIIAMS1} we list the type of branes or states and
their actions in the two $S$-dual theories, type IIA, and HW. The
branes are denoted by standard notation, $F$/$D$ stands for a
fundamental/Dirichlet brane, followed by the number of spatial
dimensions of the brane. The notation $KK$ denotes Kaluza-Klein
states. In each row the branes or $KK$ states are related by
duality, and their actions are related by (\ref{IIAMS1duality}).
\begin{table}[h]
    \begin{center}
    \begin{tabular}{||c|c|c||c|c||}
    \hline \hline
     & \ \ \ IIA \ \ \  &   & \ \ \ MS1  \ \ \  &   \\
    \hline\hline
     & $ brane $ &\ \ \  action \ \ \ & brane & \ \ \ action \ \ \ \\
   [.6ex]  \hline
 \ *\ & $ D0 $ & $ \frac{1}{g_{IIA}} \frac{R}{l_{IIA}}  $
    & $ KK $ graviton & $ \frac{R}{\rho}$ \  \\
    [.6ex] \hline \hline
   \ *\ &\ $ F1 $ & $ \left(\frac{R}{l_{IIA}}\right)^2  $
    & $ M2 $ longitudinal  & $
    \left(\frac{R}{l_{11}}\right)^2 \frac{\rho}{l_{11}} $\ \\
       [.6ex] \hline\hline
\ &\ $D2$ & $ \frac{1}{g_{IIA}} \left(\frac{R}{l_{IIA}}\right)^3$
    &\ $ M2 $ transverse & $\ \left(\frac{R}{l_{11}}\right)^3$\ \\
       [.6ex] \hline\hline
 \ &\ $ D4 $ & $ \frac{1}{g_{IIA}}
 \left(\frac{R}{l_{HE}}\right)^5  $
 & $ M5 $ longitudinal &
 $ \left(\frac{R}{l_{11}}\right)^5 \frac{\rho}{l_{11}}$\ \\
 [.6ex] \hline\hline
 \ &\ $ F5 $ & $ \frac{1}{g_{IIA}^2}
 \left(\frac{R}{l_{HE}}\right)^6  $
    &\  $ M5 $ transverse & $\ \left(\frac{R}{l_{11}}\right)^6$\ \\
 [.6ex] \hline\hline
  \end{tabular}
  \caption{BPS branes and actions in IIA/MS1. Dominant
  contributions are marked by *.}
  \label{tbIIAMS1}
\end{center}
  \end{table}

\subsection{ Type IIA vs. type IIB}

In type IIA  and type IIB theories, the 4d Yang-Mills coupling
$\alpha_{YM}= \frac{g_{YM}^2}{4 \pi}$, and Newton's constant $G_N$
are given by table \ref{tbcIIAIIB}.
\begin{table}[h]
    \begin{center}
\begin{tabular}{||c||c|c||}
    \hline \hline
     &     IIA &  IIB     \\
  [.5ex]  \hline\hline
$\ \alpha_{YM}\ $ & $ g_{IIA}^2
\left(\frac{l_{II}}{R_{IIA}}\right)^6$
     & $ g_{IIB}^2
\left(\frac{l_{II}}{R_{IIB}}\right)^6$ \\
   [.5ex]  \hline
 $ G_N $ & $\  g_{IIA}^2 \left(\frac{l_{II}}{R_{IIA}}\right)^6
 l_{IIA}^2  \ $ & $\ g_{IIB}^2
 \left(\frac{l_{II}}{R_{IIB}}\right)^6 l_{II}^2 \  $ \\
         [.5ex] \hline\hline
\end{tabular}
   \caption{ $\alpha_{YM}$ and $ G_N $ in IIA/IIB.}
  \label{tbcIIAIIB}
\end{center}
\end{table}
For one single $T$-duality the relations between the  IIA, IIB
couplings are
\begin{eqnarray} \label{IIAIIBduality}
l_{IIA} &=& l_{IIB}\equiv l_{II} \nonumber \\
 R_{IIA} &=& \frac{l_{II}^2} {R_{IIB}} \\
 g_{IIA} &=& \left( \frac{ l_{II}}{R_{IIB}}\right) g_{IIB}.
 \nonumber
\end{eqnarray}

In table \ref{tbIIAIIB} we list the type of branes or states and
their actions in the two $T$-dual theories, type IIA, and IIB. The
notation $KK$ denotes Kaluza-Klein states. In each row the branes
or $KK$ states are related by duality, either in a transverse
(denoted by $T$) or a longitudinal direction (denoted by $L$), and
their actions are related by (\ref{IIAIIBduality}).
\begin{table}[h]
    \begin{center}
    \begin{tabular}{||c|c|c||c||c|c||}
    \hline \hline
     & \ \ \ IIA \ \ \  &  & T/L & \ \ \ IIB  \ \ \  &   \\
    \hline\hline
     & $ brane $ &\ \ \  action \ \ \ & & brane & \ \ \ action \ \ \ \\
   [.6ex]  \hline
 \ *\ & $ D0 $ & $ \frac{1}{g_{IIA}} \frac{R_{IIA}}{l_{II}}  $
    & L & $D\!-\!1$& $ \frac{1}{g_{IIB}}$ \  \\
    [.6ex] \hline \hline
    \ *\ & $ D0 $ & $ \frac{1}{g_{IIA}} \frac{R}{l_{II}}$
    & T & $D1 $ & $ \frac{R R_{IIB}}{g_{IIB}l_{II}^2}$ \  \\
    [.6ex] \hline \hline
 \ * \ & $ F1 $ & $ \frac{RR_{IIA}}{l_{II}^2}$
    & L & {$KK$ MM} & $ \frac{R}{R_{IIB}}$ \  \\
    [.6ex] \hline \hline
    \ *  \ & $ F1 $ & $ \frac{R^2}{l_{II}^2}$
    & T & {$F1$ } & $ \frac{R^2}{l_{II}^2}$ \  \\
    [.6ex] \hline \hline
   \ \ & $ D2 $ & $\frac{1}{g_{IIA}} \frac{R^2R_{IIA}}{l_{II}}$
    & L & $D1$ & $ \frac{1}{g_{IIB}}
    \left(\frac{R}{l_{II}}\right)^2$ \  \\
    [.6ex] \hline \hline
 \ \ & $D2$ & $\frac{1}{g_{IIA}}
    \left(\frac{R}{l_{II}}\right)^3$
    & T & $D3$   & $\frac{1}{g_{IIB}}
    \frac{R_{IIB}R^3}{l_{II}^4}$  \  \\
    [.6ex] \hline \hline
  \end{tabular}
  \caption{BPS branes and actions in IIA/IIB. Dominant
  contributions are marked by *.}
  \label{tbIIAIIB}
\end{center}
  \end{table}
$KK$ MM stands for Kaluza-Klein momentum mode. The size of a
direction that is not T-dualized is denoted $R$. Unlike the
previous cases of S-duality, we have to make a distinction between
different radii, since even if their sizes were equal to begin
with, the T-duality changes them.

This completes the basic relations between the theories at the
edges of the star diagram \ref{fig1}. There are additional
relations between the perturbative string theories, but they are
generated by the relations we have listed.

\subsection{Discussion}

Our  conjecture has been  that the all the possible string
non-perturbative effects are accounted for by the D and F
instantons (in 4D) given in the above tables. We note that all
expected SNP  in each theory appear, and that there's no room for
exotic SNP. For example, suppose that in the HO theory we look for
SNP of strength $e^{- {\frac{1}{g_{HO}}
\left(\frac{R}{l_{HO}}\right)^\beta}}$. as  might be required for
Kahler stabilization \cite{bd}. By S-duality this would require an
effect in type I of the form $e^{{-g_{I}^{1-\beta}
\left(\frac{R}{l_{I}}\right)^{2\beta}}}$, which surely cannot
exist. Similarly we would like to argue that in the HW theory the
correction to the Kahler potential coming from  transverse
membranes \cite{choikim} actually vanishes since (see table
\ref{tbHEHW}) and the related discussion) this has no analog in
the S-dual HE theory. Indeed it would be surprising (and contrary
to the spirit of String Universality) if such an NP contribution
to the HW theory were present since there appears to be no way
this could arise in the HO theory or its S-dual type I theory.
Another piece of supporting evidence for this is that in the
IIA/MS1 case (see table \ref{tbIIAMS1}) all  contributions in the
strong coupling theory have a weak coupling analog so it would be
strange if this were not the case in the HW theory.

Considering only the leading exponential dependence it is clear
that moduli are unstable, since they have runaway potentials which
force them towards free 10d (11d) theories. We therefore conclude
that moduli stabilization cannot occur when either the inverse
coupling or the volume are parametrically large.

\section{Moduli stabilization and SUSY breaking}

In this section we will discuss possible mechanisms for moduli
stabilization, and their likely values. We have already argued
that stabilization of moduli in the  boundary region of moduli
space is unlikely. Here we explicitly show why this is so. We then
argue that stabilization near the self-dual point, motivated by
string universality, is plausible, and that SUSY breaking is
likely to occur at a much lower scale than moduli stabilization
scale. To discuss moduli stabilization in more detail we need to
take into account supersymmetry in a more quantitative way. Moduli
are chiral superfields of $D=4$, $N=1$ SUGRA, which means that
their interactions are constrained by the general form of the
$N=1$ SUGRA Lagrangian.

The potential for moduli $\phi_i$ in $N=1$ supergravity is given
by \cite{cremmer}\footnote{A convenient source for the formulae in
this section is volume II chapter 18 of \cite{jp} which should be
consulted for the original references. See also the reviews
\cite{nilles}, \cite{ovrut}.} the following expression\footnote{In
this section we will put $\k=M_P^{-1}=1$ for convenience.}
 \be\label{potential}
V=e^{\cG}(\cG_i\cG^{i\bar j}{\bar\cG_j}-3)= F_i{\cG}^{i\bar
j}\bar{F_j}-3e^{\cG},
 \ee
and $\cG$ is usually written as
 \be
\cG=K(\phi_i,\bar\phi_i)+\ln{|W(\phi_i)|^2},
 \ee
where the real analytic function $K$ is the Kahler potential, the
holomorphic function $W$ is the superpotential and $F_i\equiv
e^{\cG\over 2}\cG_i$ is an order parameter for SUSY breaking. Also
 \be
\cG^{i\bar j}=\cG^{-1}_{i\bar j},\cG_i={\pa\cG\over\pa\phi_i}
~\cG_{i\bar j}={\pa^2\cG\over\pa\phi_i\phi_{\bar j}}.
 \ee
The potential $V$ can be expressed in terms of $K$ and $W$ as
follows,
 \begin{equation}
V=e^{K}\left[D_iWK^{i\bar j}\bar{D_j W}-3|W|^2\right]
 \ee
where $D_iW=\pa_iW+\pa_iKW$. The coupling to the gauge sector
modifies the $F$-terms into
 \be\label{F}
 F_i=e^{\cG\over 2}\cG_i+\frac{1}{4}f_i\l\l,
 \ee
(where $f_i= {\pa f\over\pa\phi^i}$,  $f$ being the gauge coupling
function) and the potential can then still be written in the form
given in the second equality of (\ref{potential}).

Let us consider string theory compactified on a 6-manifold which
preserves $N=1$ SUSY. Conventionally a four dimensional complex
modulus $S$ (related to the model independent 10D dilaton/axion)
and three complex moduli $T_i$ (related to the size and shape of
the compact manifold) are defined. In particular, we have
 \be
\label{ST} Re S=e^{-2\phi}\sqrt{\det G_c} \ee where $\phi$ is the
10D dilaton and the last factor on the RHS is  the measure on the
compact manifold. Then in perturbative string theory we have at
tree level,
 \be\label{pertG}
\cG = -\sum_{i=1}^{3}\ln (T_i+\bar T_i-C^a\bar{C^a})-
 \ln (S+\bar S)+\ln |W(C)|^2,
 \ee
where $C^a$ are matter fields coming from the 10D gauge fields
tangent to the compact directions. The first term comes from the
metric of the 6 manifold (CY space or orbifold)  which is
parametrized in terms of the three complex $T_i$ moduli. This form
of the function $\cG$ is (a generalization of ) the so-called
no-scale model \cite{noscale} which leads to a positive definite
potential. Up to field redefinitions (involving the S and T
fields) it is also obtained from the compactification of the HW
theory \cite{nilles,ovrut}.

We would like at this point to emphasize the appearance and
possible implications of the ``practical cosmological problem",
which as we explained is not the same as the ``cosmological
constant problem", whose solution should not be a requirement of
phenomenological SUSY breaking models.  In this particular context
the ``practical cosmological problem" takes the following form. To
allow models of supersymmetry breaking to predict reliably the
structure of soft-supersymmetry-breaking terms it is essential
that the absolute value of the potential at its minimum does not
exceed\footnote{Assuming that there are no additional phase
transitions at intermediate scales in between the SUSY breaking
scale and the scale at which it becomes explicit.} the order of
$m_{3/2}^4$, where $m_{3/2}$ is the gravitino mass which
determines  a typical size of soft breaking terms. In most
supergravity models (not of the no-scale type) the value of the
potential at the minimum has a tendency to be negative and of
order $M_P^2 m_{3/2}^2$. In addition to the loss of predictive
power this feature has obvious disastrous cosmological
consequences, to be avoided.  Most non-perturbative SUSY breaking
models suffer from this problem and special measures have to be
taken to remedy the situation.

We discuss three different classes of models:
\begin{itemize}
\item
no-scale type models, which solve automatically the practical
cosmological constant problem (at least at tree level), but have
trouble in stabilizing moduli and predicting non-vanishing gaugino
masses.
\item
``race-track" type models which have problems breaking SUSY,
stabilizing moduli and solving the practical cosmological constant
problem.
\item
our suggestion for stabilization around the self-dual point.
\end{itemize}

\subsection{No scale models}

No scale models are models for which (\ref{pertG})
holds\footnote{The discussion in this subsection is included for
reasons of clarity and completeness.  All the material given here
can be found in the literature.}. Let us assume now for simplicity
that the three $T$ moduli are identified. At string tree level the
gauge coupling function $f=f_SS$ so that,
$F_T=e^{\shalf\cG}\cG_T,~F_C=e^{\shalf\cG}\cG_C,~F_S=
e^{\shalf\cG}\cG_S+{1\over 4}f_S(\l\l )$ and the potential
becomes,
 \begin{equation}
V=F_s\cG^{S\bar S} \bar{F_S}+e^{\tilde K}{|\pa_CW|^2
 \over 3(T+\bar T-|C|^2)^2}.
 \ee
This is a positive definite potential whose minimum is at $F_S=0$
and $\pa_CW=0$. SUSY may still be broken if $F_T\ne 0$ and/or
$F_C\ne 0$. This requires that the superpotential at the minimum
be non-vanishing. In the tree level compactification
$W=d_{ijk}C^iC^jC^k$ where $d_{ijk}$ are coupling constants. At
the minimum we must have
$\pa_lW=d_{ijl}C^iC^j+d_{ilj}C^iC^j+d_{lij}C^iC^j=0$, but this
implies that $W=0$ at the minimum and no SUSY breaking. So to have
broken SUSY in this scenario one must require a constant in the
superpotential $W_0=\L^3$ (generated by some yet unknown
non-perturbative mechanism). Furthermore $\L$ must be around
$10^{13}GeV$ in order to get the required scale of SUSY breaking.
It is not at all clear why such a scale should arise but if it
does exist\footnote{In available examples constants in the
superpotential  are quantized in units of the string scale.} then
(taking into account gaugino condensation) the dilaton is
stabilized with SUSY broken at an acceptable scale.

As is well known, no scale type models have several problems even
if one assumed that the form of the $\cG$ and the tree level
result for $f$ survived string loop corrections and that there is
a constant in the superpotential resulting in SUSY breaking f the
desired magnitude. We list them here for completeness.
\begin{itemize}
\item
 a) The T-moduli are not determined.
\item
 b) Since SUSY breaking is not
dilaton dominated generically there will be flavor changing
neutral currents.
\item
 c) The gaugino masses are given by the formula
$m_{\l}=<f_i\cG^{i\bar j}\bar{F_j}>$ \cite{cremmer}. If $f=f_SS$
as in tree level string theory $f_i=f_S$ and since $F_S=0$ at the
minimum this mass vanishes (note that $G^{S\bar T}=0$ in these
models).
\end{itemize}

Of course the gauge coupling function is corrected by string loop
effects and one should write $f=f_SS+f_TT$. But in this case there
are two additional terms in the potential. i.e. defining
$Q_i={1\over 4}f_i\l\l$
 \begin{equation}
Q_T\cG^{T\bar T}Q_T+2ReQ_T\cG^{T\bar T}e^{\shalf\cG}\cG_T.
 \ee
The second term clearly spoils the positive definiteness of
the potential. Alternatively if one works with an effective
Lagrangian after gaugino condensation one has a superpotential
that is now T-dependent and that will spoil the no-scale property.
Thus even if one assumed that the general no-scale form of the
Kahler potential (\ref{pertG}) is unaffected by quantum
corrections one cannot preserve the positive definiteness of the
potential and get non-zero gaugino masses.

Since the no-scale field is expected to be quite light, and its
interaction are of gravitational strength, a particularly severe
potential problem for this class of models is the amount of energy
stored in the no-scale scalar field and the implications of this
energy on late cosmological evolution. This is a manifestation of
the so-called moduli problem  \cite{moduli1,moduli2}

\subsection{``Race track" models: Stabilization near the
 boundary of moduli space}

An alternative to no-scale models is that the parameters of the
superpotential and the Kahler potential are chosen in such a way
that results in a SUSY breaking minimum with  vanishing
cosmological constant. This has been the class of models of choice
for most previous works on string phenomenology, and it was
recently revived in a slightly different context
\cite{dineshirman}. It is usually assumed that stabilization and
SUSY breaking occur at the same scale, which has to be much below
the string scale. This leads to known problems, which we recall
for emphasis, and further argue that they are generic to such
models.

We will argue shortly that it is very unlikely that a minimum of
vanishing cosmological constant which breaks SUSY is found at weak
coupling and large compactification volume (which we called ``the
boundary region of moduli space"). But even if we assume that such
a minimum exists then there is always a deep minimum with a  large
negative cosmological constant towards weaker coupling and larger
volume \cite{bs}. In addition, there's always a supersymmetric
minimum at vanishing coupling and infinite volume. This
multi-minima structure brings into focus the barriers separating
them.  If these barriers are high enough one may argue that flat
space is a metastable state with a large enough life time.
Generically, however, this is not the case, and classical or
quantum transitions between minima are quite fast. In the context
of gaugino-condensation race-track models this was discussed in
\cite{bs}.  In particular, in a cosmological setup it was shown
\cite{bs} that classical roll-over of moduli towards weak coupling
and large volume are generic, and occur for a large class of
moduli initial conditions. Later it was shown that cosmic friction
can somewhat improve the situation \cite{copeland}, and recently
it was argued that finite temperature effects drastically improve
the situation \cite{hsow}.

We would like to show that the problems arising in this class of
models are rooted at their basic assumptions, and that they cannot
be remedied by choosing different parameters or playing with
numbers. Our conclusion is that, at the very least, it is
inconsistent to consider, in this class of models, only $T$ and
$S$ moduli, and parametrize SUSY breaking  in terms of a complex
vector in $(F_T,F_S)$ plane as suggested in \cite{kaplouis,bim}.
This conclusion was in fact already recognized in
gaugino-condensation models by \cite{bim}, where an additional
ad-hoc chiral superfield was included, and later also in
\cite{moriond,casas}.

Let us now examine more closely the possibility of stabilizing
moduli near the  boundaries of moduli space. We consider generic
moduli chiral superfield , which we denote by $S$, which could be
either the dilaton $S$-modulus, or the $T$-modulus. We assume that
its Kahler potential is given by $K=-\ln (S+S^*)$, and that $Re
S>0$, corresponding to having a well defined compactification
volume and gauge coupling. The generic feature of the
superpotential $W(S)$ near the boundaries of moduli space is its
steepness. This has to be so, because we insist that a new scale,
much lower than the string scale is generated dynamically, and the
ratio of this new scale to the Planck scale has to be reached
within about Planck distance in moduli space. This requires that
derivatives of the superpotential are large. In mathematical
terms, the steepness property of the superpotential is expressed
as follows,
 \bq
 \frac{\left|(S+S^*)\partial_S^{(n+1)} W\right|}
 {\left| \partial_S^n W \right|}\gg 1\ \ \
 n=0,1,2,3.
  \label{a20}
  \eq
This property certainly holds for all ``gaugino-condensation"
superpotentials, but as explained, it is generic to all models of
stabilization around the  boundaries of moduli space. The typical
example of a superpotential satisfying (\ref{a20}) is a sum of
exponentials $W(S)=\sum_i e^{-\beta_i S}$, with $Re \beta_i \gg
1$, in the region $|S\beta_i| >> 1$. In this example the
``boundary region of moduli space" is simply the region $Re S \gaq
1$, but in general, the precise definition will depend on the
details of the model. It is good to keep this example in mind
while going through the following arguments, but we will not use
any particular specific form for $W$.

Inequality (\ref{a20}) holds as a functional relation and can be,
of course,  violated at isolated points. Obviously (see eq.
(\ref{a6})), it is violated at extrema. But the violations of
(\ref{a20}) are only in some of the relations between the
derivatives, for example the first and second derivatives, while
for the rest, the rule that the higher the derivative, the larger
it is, still holds.

The potential $V$, is given in terms of the superpotential $W$,
and its derivatives
 \bq
V(S,S^*)=(S+S^*) F(S,S^*) F^*(S,S^*)- \frac{3}{S+S^*} W(S)
W^*(S^*),
 \label{a1}
 \eq
where $ F(S,S^*)=\der_S W(S)-\frac{1}{S+S^*} W(S).$
The first derivatives of the potential are give by
 \brr
\der_S V&=&(S+S^*) \der^2_S W(S) F^*-\frac{2}{S+S^*} F W^*\nnu\\
\der_{S^*} V&=&(S+S^*) \der^2_{S^*} W^*(S^*) F^*-\frac{2}{S+S^*}
F^* W.
 \label{a5}
 \err
An extremum is determined by solutions of $\der_S V=0$, that is,
 \bq (S+S^*)^2 \der^2_S W(S) F^*=2 F W^*,
  \label{a6}
   \eq
 which can be satisfied in two ways,
 \brr
  F&\ne&0\ \ {\rm
and}\hspace{0.3in} (S+S^*)^2 \der^2_S W(S) F^*=2 F W^*\\
 F&=&0.
\label{a7}
 \err
At an extremum with a vanishing $F$ as in (\ref{a7}), SUSY is
unbroken and the cosmological constant is $-3|W|^2$, which is
generically too large to obey our requirement that it solves the
practical cosmological constant problem. If $W$ is also tuned to
zero at the minimum, then one encounters the problems associated
with the appearance of an additional deep minimum with negative
cosmological constant and those arising from the multi-minima
structure which were alluded to previously. Only an extremum as in
(\ref{a6}) breaks SUSY, but as we will show shortly it is never a
minimum for steep potentials.

To determine whether  the extrema are minima, maxima or saddle
points we need to analyze the matrix of second derivatives of the
potential at the extrema. Using the following expressions for the
derivatives of $F$
 \bq
 \der_S F(S,S^*)=\der^2_S W(S)-\frac{1}{S+S^*} F(S),
 \eq
 and
 \bq
 \der_{S^*} F(S,S^*)=\frac{1}{(S+S^*)^2} W(S),
 \label{a4}
 \eq
and similar expressions for derivatives of $F^*$, we can compute
second derivatives of $V(S,S^*)$,
 \bq
 \der^2_{SS^*} V= -\frac{2}{(S+S^*)^3} W W^*
 -\frac{2}{(S+S^*)} F F^* + (S+S^*) \der_S^2 W \der_{S^*}^2 W^*
 \label{a8}
 \eq
 \bq
\der^2_S V= \frac{4}{(S+S^*)^2} F W^* -\frac{1}{S+S^*}
 \der_S^2 W W^*+\der^2_S W F^* + (S+S^*) \der_S^3 W F^*.
  \label{a9}
  \eq
Expressions (\ref{a8},\ref{a9}) take the following form,\\
 Case (i): $F=0$; $\der_S V=0$ at the extremum $S_0$, then
at $S_0$
 \bq
 \der^2_{SS^*}V=-\frac{2}{(S+S^*)^3} W W^* + (S+S^*)
 \der_S^2 W \der_{S^*}^2 W^*,
 \label{a10}
 \eq
 and
 \bq
 \der^2_S V=-\frac{1}{S+S^*} \der_S^2 W W^*.
 \label{a11}
 \eq
Case (ii): $F\ne 0$; $\der_S V=0$ at the extremum $S_0$, then
at $S_0$
 \bq
 |(S+S^*)^2 \der^2_S W|=|2 W|,
 \label{a12}
 \eq
 and therefore
 \bq
 \der^2_{SS^*}V=\frac{2}{(S+S^*)^3} W W^*
 - \frac{2}{S+S^*} F F^*,
 \label{a13}
  \eq
  and
  \bq
  \der^2_S V= (S+S^*) \der_S^3 W
F^*-\frac{1}{S+S^*} \der_S^2 W W^* +\frac{6}{(S+S^*)^2} F W^*.
 \label{a14}
 \eq
{}From these expressions we can calculate the various partial
derivatives with respect to $S_R= Re(S)$ and $S_I=Im(S)$. Using
 \brr
 \frac{\der^2 V}{\der S_R \der S_R}&=&
 +\frac{\der^2 V}{\der S \der S}
 +\frac{\der^2 V}{\der S^* \der S^*} +
 2 \frac{\der^2 V}{\der S \der S^*} \nnu \\
  \frac{\der^2 V}{\der S_I \der S_I}&=& -
  \frac{\der^2 V}{\der S \der S} -
 \frac{\der^2 V}{\der S^* \der S^*} +
 2 \frac{\der^2 V}{\der S \der S^*} \nnu \\
 \frac{\der^2 V}{\der S_R \der S_I}&=& +
 \frac{\der^2 V}{\der S \der S} -
 \frac{\der^2 V}{\der S^* \der S^*}.
 \label{a15}
 \err
The relevant quantity is the determinant of the matrix of second
derivatives,
 \brr
 H &=&\frac{\der^2 V}{\der S_R
 \der S_R} \frac{\der^2 V}{\der S_I
\der S_I}
 - \left(\frac{\der^2
V}{\der S_R \der S_I}\right)^2 \nnu \\
 &=&-4 \left( \der_{SS}V \der_{S^*S^*} V -\der_{SS^*}V \der_{S^*S}
V\right).
 \label{a16}
 \err

We can now substitute Eqs. (\ref{a10}, \ref{a11}) and (\ref{a13},
\ref{a14}) into (\ref{a16}) and obtain expressions for $H$ at the
extrema. First, Case (i),
  \bq
  H=4\left( (S+S^*)^2 (|\der^2_S W|^2)^2
  -\frac{5}{(S+S^*)^2} |\der^2_S W|^2 |W|^2
  +\frac{4}{(S+S^*)^6} (|W|^2)^2\right),
  \label{a17}\eq
where we have used eqs.(\ref{a10}, \ref{a11}). Expression
(\ref{a17}) for $H$ can be written as
 \bq
 H=4\left( (S+S^*) |\der^2_S W|^2
 -\frac{2}{(S+S^*)^3} |W|^2\right)^2
 -\frac{4}{(S+S^*)^2} |\der^2_S W|^2 |W|^2.
 \label{a18}
 \eq
This means that the extremum of type (i) is either a local minimum
or a local maximum. To check which of the two, it is enough to
choose an arbitrary direction and check if the second derivative
is positive or negative. For example, choose the $S_R$ direction.
{}From eq.(\ref{a15}) we obtain
 \bq
\der^2_{S_R S_R} V= 2(S+S^*) |\der^2_SW|^2- \frac{2}{S+S^*}
 Re(\der^2_SW W^*)- \frac{4}{(S+S^*)^3}|W|^2. \label{a19}
 \eq

Using Eq.(\ref{a20}), we see that if $\der_S^2 W\ne 0$ at the
extremum of case (i), then from (\ref{a19})
 \bq
 \der^2_{S_R S_R} V\approx 2(S+S^*) |\der^2_SW|^2>0.
 \eq
The conclusion is that a generic extremum of type (i) is  a
minimum. If additional conditions are imposed on the
superpotential it may be possible to have a local maximum at an
extremum of type (i).

The analysis in case (ii) is a little bit more complicated but is
essentially the same analysis. For case (ii), rather than write
the full complicated expressions, we analyze them using
eq.(\ref{a20}). If $\der_S^3 W\ne 0$ at the extremum then
 \bq
 H\approx -4 (S+S^*)^2 |\der_S^3 W|^2 |F|^2,
  \label{a21}
  \eq
so $H<0$ and the extremum is necessarily a saddle point. If
$\der_S^3 W=0$ at the extremum, then using Eq.(\ref{a12}), one
finds that $H\approx +\frac{16}{(S+S^*)^2}(|F|^2)^2$ so $H>0$ and
the extremum is either a maximum or a minimum. Checking, for
example, $\der^2_{S_R S_R} V\approx -\frac{4}{S+S^*}|F|^2$ reveals
that this is a local maximum. No further simplification can occur
and therefore the analysis is complete. The conclusion is that,
under the conditions of Eq.(\ref{a20}),  an extremum of case (ii)
is either a saddle point or a maximum, but never a minimum.

So far we have used the moduli perturbative Kahler potential
$K=-\ln(S+S^*)$ in our discussion. We expect the results of our
analysis to be similar if this Kahler potential
receives some moderate corrections. If, as expected,  corrections
to the Kahler potential preserve the steepness of the potential
so that superpotential derivatives are larger than derivatives of the Kahler
potential, an analog of (\ref{a20}) exists, in which
powers of $(S+S^*)$ are replaced, where appropriate,
by $e^K$, $\der_S K$ or $\der_{S^*} K$.
The subsequent analysis follows through, since
the essential ingredient that we have used was a classification
of the largest terms in the equations, which should still be valid.
We do not consider the case when corrections to
the Kahler potential are larger than its original perturbative value, since this
would mean that perturbation theory is badly broken in the outer region
of moduli space, an unlikely situation in contradiction with
available information.

To summarize, we have found that it is not possible to find a SUSY
breaking minimum in the region where the superpotential is a steep
function. This is a very general conclusion which shows how hard
it is to make ``race-track" models work. This conclusion has been
reached previously using different arguments in the context of
gaugino condensation models \cite{bs,bim,casas}.

\subsection{Stabilization around the self-dual point}

We propose that moduli stabilization around the self-dual point is
a plausible scenario, allowing the possibility of SUSY breaking
while evading the problems we discovered for the other scenarios.
Our discussion will be qualitative, postponing the quantitative
analysis to a dedicated project \cite{future}.

Our suggestion, motivated by string universality is the following:
\begin{itemize}
\item
Moduli are stabilized at a scale below, but not much below, the 4D
Planck scale, which is similar to the string scale. All flat
directions are lifted at this stage, leaving no light moduli. A
possible source of moduli potential can be SNP. SUSY is unbroken
at that scale, and the cosmological constant at the minimum  is
parametrically smaller than $M_P^4$. The dynamics which results in
this must be intrinsically stringy.

\item
SUSY is broken at a much lower intermediate scale, by additional
small non-perturbative effects, which do not spoil the
stabilization of moduli. The dynamics here probably can  be
understood in field theoretic terms, for example
gaugino-condensation in the hidden sector.
\end{itemize}
The moduli stabilizing superpotential should be of order $M_P^3$,
and all the F-terms, and the superpotential itself should vanish
at the minimum. SNP are likely to obey these requirements, which
are nothing but the consistency requirement that flat space is
stable to SNP. Since BPS branes are solutions of string/M-theory
in a flat background, there is no reason to expect that they will
destabilize flat-space.

The SUSY breaking superpotential should be of order
 \begin{equation}
 \label{delW}
 \delta W\sim m_{3/2} M_P^2,
\end{equation}
and the resulting $F$ term is of order
\begin{equation}
 \label{delF}
 \delta F\sim m_{3/2} M_P,
\end{equation}
which is not expected to destabilize a minimum with curvature of
the order of $M_P^2$, since the new location of the minimum is
shifted by a small amount proportional to $m_{3/2}$.

We argue that the scenario that we are proposing will not suffer
from the problems that the other scenarios were inflicted with.
First, a minimum with broken SUSY as in case (ii) (eq.\ref{a12})
can be found, since the arguments against such a possibility
depended on the steepness of the superpotential, but in the
current situation the superpotential is not a steep function. The
practical cosmological constant problem can be solved provided the
superpotential satisfies the following additional condition,
 \begin{equation}
 \label{goodmin}
 (S+S^*)^2 |F|^2= 3|W|^2/M_P^2.
 \end{equation}
That eq.(\ref{goodmin}) can be satisfied is clear from comparing
eqs. (\ref{delW}) and (\ref{delF}). The problems associated with
the multi-minima structure become as benign as possible, since the
height of the barrier between the SUSY breaking minimum around the
self-dual point and the SUSY preserving minimum at infinity is of
order $M_P^4$ and its width is of order $M_P$, which are the best
values that we can hope for.

The above arguments are not a proof that our proposed scenario
works, but they show that it is plausible, and that it may not
suffer from the problems that existing ideas for moduli
stabilization and SUSY breaking.

Previously, Kaplunovsky and Louis \cite{kaplouis2} have proposed
in the context of $F$-theory a scenario which has some of the
ingredients that we are proposing, namely, stabilization at a high
scale and SUSY breaking at a lower scale. However, as pointed out
in \cite{dine} their proposal suffered from drawback that a field
theory argument is inconsistent at the string/Planck scale. What
we are proposing on the other hand is that the moduli are
stabilized at the string scale by intrinsically stringy effects
while the SUSY breaking could be field theoretic, in the spirit of
\cite{dns}.

\section{The consistency of string universality with phenomenology}

In this section we would like to show that if indeed moduli are
stabilized near the self-dual point, then the phenomenology of the
effective low energy theories derived from  the various string
theories is consistent with each other and with expectations.  Of
course, this  analysis is done using perturbative theories in the
boundary region of moduli space, but we hope to demonstrate
that approaching the central region from different directions
gives a consistent picture of the physics there.

\subsection{The value of $\rho$ in Horava-Witten Theory}

In string theories I, HO the point $g =1$ is the one where one
would expect perturbation theory to break down. The phenomenology
of these theories is not inconsistent with this value (assuming
that the perturbative phenomenology can be extrapolated with
suitable assumptions about stabilization of moduli). In
particular, with threshold corrections there  is no inconsistency
with the value of the Planck mass.  On the other hand the HW
phenomenology would seem to yield a large value of the eleventh
dimension and hence a large value of the string coupling. Let us
therefore examine this question in some detail.

In eq.(\ref{HWea}) we take ${\cal M}_{11} =R^{10}\times S^1$ and
work with fields that are reflection symmetric under
$x^{11}\rightarrow -x^{11}$. In the zero'th order calculation one
assumes that $M^{11}$ can be compactified as a direct product
space $R^4\times CY_3\times S^1$. Defining the volume of CY
$\int\sqrt{G_6}d^6x= V$ and putting $\int dx^{11}
\sqrt{G_{11}}=2\pi\rho$ as before we may identify,
 \be\label{couplings}
G_N={\k_{11}^2\over
16\pi^2V\rho},~~~~\a_{GUT}={(4\pi\k_{11}^2)^{2/3}\over 2V}.
 \ee
If we identify $V={a^6\over M_{GUT}^6}$ where $a\sim O(1)$, then
we obtain,
 \begin{eqnarray}
\label{couplingsB1}
 \rho&=& a^3{(2\a_{GUT})^{3/2}\over 64\pi^3}{M^2_{P} \over
 M^3_{GUT}} \sim a^3{1.8\over M_{GUT}} \\
\label{couplingsB2}
 \k_{11}^{2\over 9}&=&{(2V\a_{GUT})^{1/6}\over (4\pi )^{1\over 9}}=
 0.5{a\over M_{GUT}}.
 \end{eqnarray}
In the above $M_P^2=G_N^{-1}=1.2\times 10^{19}GeV$ and we have put
$\a_{GUT}=1/25,~~M_{GUT}=3\times 10^{16}GeV$. For the low energy
M-theory expansion to make sense we should have \cite{ew},
${\k_{11}^{4/3}\over V} <1$ and ${\rho(\k_{11})^{2/3}\over
V^{2/3}}<1$. With the above numbers the former is about 0.02 but
the latter is $0.3a^2$ so that we must have $a\sim O(1)$ and seems
to rule out values close to $2\pi$. We define the eleven
dimensional Planck length  by the relation, $2\k_{11}^2=(2\pi
)^8l_{11}^9$ as before. This and the radius of the M theory circle
are related to the HE string length and coupling constant by
eq. (\ref{HWHE}). So we may summarize this naive comparisons
of scales as follows,
 \begin{eqnarray}
 \label{scales1}
 {\pi\rho\over 2\pi l_{11}}& \simeq & 9a^2 \\
 \label{scales2}
 {\pi\rho\over V^{1/6}}& \simeq & 6 a^2 \\
 \label{scales3}
 {V^{1/6}\over 2\pi l_{11}}& \simeq & 1.6.
 \end{eqnarray}

Thus we see that with $a\sim 1$, the extra (eleventh) dimension is
about an order of magnitude larger than the other length scales of
the theory. However it should be noted that although consistency
of the arguments require that $a$ has the upper bound given above,
there is no lower bound. Indeed the last ratio is independent of
$a$ but larger than one as required.

In order to compare with perturbative HE string theory (see
\ref{HEHWduality}) we make the identification $\rho
=g^{2/3}_{HE}l_{HE}$. Then we get from the above $g_{HE}=76a^3$
which must certainly be considered a value that is in the strong
coupling region if $a\sim 1$. However, $a$ is cubed in this
relation and with $a\simeq 0.25$ we would obtain $g\sim O(1)$.  It
is not unreasonable to expect a numerical coefficient of this
magnitude in the relation between the observed unification scale
and the Kaluza-Klein scale, but we will presently argue that if
one also takes into account threshold effects one can indeed get
values of $g\sim 1$ even with $a \sim 1$.

If the phenomenology does indeed require  large values of
$\pi\rho$ or equivalently of $g$, then the idea of string
universality is not viable. But then we would have to say that the
HE/HW theory is special and is the only one that gives the low
energy energy physics of the real world. The evolution of the
physical couplings would be as discussed in \cite{bd} and
illustrated in Fig 18.2 of \cite{jp} according to which a fifth
dimension opens up a little above $10^{15}GeV$ so that the
dimensionless gravitational coupling constant starts evolving as
in a 5D theory to meet the other three couplings at the GUT scale.
So as one increases the energy the world appears to go from being
4D to 5D and finally to 11D. Such a picture clearly cannot hold in
any other string theory where one should only see a transition
from 4D to 10D at the string scale of around $10^{18}$. In this
picture the dimensionless gravitational coupling does not meet the
other three at the GUT scale, one merely has a determination of
its value at that scale.

Both these scenarios cannot be true, and one has to pick one over
the other. Such a situation seems very strange to us. It would
mean that the other string theories have no role to play in
nature. The existence of duality relation among them and with the
HE theory however seems to argue against such an interpretation.
It seems much more natural to have a situation of string
universality as discussed in the introduction. Let us therefore
reexamine the phenomenology of the HE/HW theory. One possibility
was mentioned above, i.e. one may have $a<1$. Below we will argue
that if threshold corrections are included in the relations
(\ref{couplings}) then it is possible (for non-standard
embeddings) to obtain a coupling $g\sim O(1)$ so that in the
equivalent HW picture one has a eleventh (fifth) dimension whose
size is of the order of the string length $l_{11}\sim l_{HE}$.

In the string theory this is of the order of stringy fluctuations
and does not have the interpretation of an extra dimension. In
other words the picture that emerges is that of an intermediate
coupling HE/HW theory that lies at the boundary of the  region
accessible to both. The coupling unification picture that will
emerge from this analysis will be the same as that which comes out
of any of the other phenomenologically viable string
constructions.

It is easiest to discuss the threshold corrections from the strong
coupling HW end following Witten's arguments \cite{ew} (for
reviews of work done since this original paper see
\cite{nilles,ovrut,munoz}). It should however be stressed that
this is completely equivalent to the weak coupling string
calculation extrapolated to strong (or at least intermediate)
coupling, and for large compactification volume. \cite{nilles}

As pointed out in \cite{ew} the volume of the CY space depends on
the value of $x^{11}$. Thus we have
 \begin{eqnarray}
 \label{volvar} V_H&\equiv & V(\pi\rho ) \nonumber \\
  &=& V_O+2\pi^2\rho\left
({\k\over 4\pi}\right )^{2/3} \int_{X_O}\o\w (\tr F\w F-\shalf\tr
R\w R) \nn V_O&\equiv&V(0).
 \end{eqnarray}
Calling one of the walls the observable one with CY volume $V_O$
and the other one the hidden wall with CY volume $V_H$ we may
write
 \begin{equation}
V_{O,H}=V(1\mp\e ),
 \ee
where the threshold correction $\e$, is given by the integral over
the CY space $X_O$ at the observable wall,
 \be
\e={2\pi^2\rho\over 2V}\left ({\k\over 4\pi}\right )^{2/3}{1\over
8\pi^2}\int_{X_{O}}\o\w (\tr F\w F-\shalf\tr R\w R).
 \ee
and $V=<V>={V_O+V_H\over 2}$. $\e$ is negative for the
standard embedding but may be positive or negative for
non-standard embeddings. Also the requirement that $V_{O,H},V>0$
implies that $|\e|<1$.

Now we identify $\a_O=\a_{GUT}\simeq {1\over 25}$ and
$V_O^{1/6}={a\over M_{GUT}}$. But the important point is that the
expression for Newton's constant involves the average volume V.
Redoing the calculations leading to
(\ref{couplingsB1},\ref{couplingsB2}) by taking into account this
variation of the volume of the CY space, we get
\begin{eqnarray}
\r &=& {(2\a_OV_O)^{3/2}\over 64\pi^3G_NV_O}(1-\e) =
 a^3 1.8 M_{GUT}^{-1}(1-\e ) \nn
 k_{11}^{2/9}&=&{(2V_O\a_O)^{1/6}\over(4\pi)^{1/9}}=
 0.5{a\over M_{GUT}}.
 \end{eqnarray}
Then the relations of (\ref{scales1},\ref{scales2}) acquire a
factor $(1-\e )$ on their right hand sides so that even if $a=1$,
for a non-standard embedding with $(1-\e_O)\sim O(10^{-1})$ we
would get an eleventh dimension whose size is of the order of the
eleven dimensional Planck scale (or the string scale) and hence in
the ten dimensional theory a coupling $g$ of order unity.

We should stress again that this calculation was by no means an
attempt to show that $g_{HE}=1$. It is merely an argument to
demonstrate that the conclusion that  $g_{HE}>>1$ is unwarranted.
After all we have  been arguing that along with $g=1$ the $T$
moduli should be stabilized close to the string scale (which is
the same as $l_{11}$ if $g=1$), but the above argument was a large
volume one. This is in the spirit of our whole discussion where we
approach the
 the ``middle of moduli space" from different (computable) directions to get hints on the nature of this region.

\subsection{Type I/IIB orientifold compactifications and Brane
worlds}

Recently there has been much excitement about the possibility that
the string scale is close to the weak scale ($\sim$ 1 TeV)
\cite{jl,AASD}. Within the space of possible string theoretic
formal constructs the sort of situation envisaged in \cite{jl} may
be modeled by compactifying type I string theory on a 6 torus (or
orbifold) and T-dualizing in all 6  compact directions. The
resulting theory is a type IIB orientifold with $2^6$ orientifold
planes and 32 D3 branes.

The compactified type I theory has the following 4d terms,
 \begin{equation}
 \Gamma_I = {V_6\over (2\pi )^7g^2l_I^8}\int d^4x\sqrt{G_4R}
+{V_6\over 4(2\pi )^7gl_I^6}\int d^4x\sqrt{G_4}\tr F^2.
 \ee
In this picture we write the compactification volume as $V_6=(2\pi
R)^6$ and the theory has Kaluza-Klein (KK) modes and winding modes
with masses $M_{KK}={n\over R}~n\e\cal Z$ and $M_w={wR\over
l_I^2}~w\e\cal Z$ respectively. For simplicity we have taken all
radii equal and we have only kept the constant mode of the
dilaton.

The T-dual effective action is obtained by the transformations
$R\rightarrow R'={l_I^2\over R}$ and  $g\ra g'
 = \left ({l_I\over R}\right )^6g$. We get
\begin{equation}\label{tdualI}
 \Gamma'={V_6'\over (2\pi )^7l_I^8g'^2}\int d^4x\sqrt{G} R+
{1\over 2\pi g'}\int d^4x{1\over 4}\tr F^2.
 \ee
It should be noted that the second term is just a 4 dimensional
integral because the D9 brane of type I has been transformed into
a D3 brane and $V'=(2\pi R')^6$. The two pictures above must be
physically equivalent as {\it string theories} though not
necessarily as low energy field theories. For instance even though
in the second ($D_3$ brane) picture the gauge theory modes are
just four dimensional while in the first picture they are ten
dimensional it should be recalled that the KK modes in the first
picture get replaced by winding modes in the second.

{}From the type I picture we have
\begin{equation}
G_N={1\over 8}{l_I^8g^2\over R^6}, ~~\a_{YM}=\shalf{l_I^6\over
R^6}g.
 \ee
Now if the  string scale is as low as $1\ TeV$ then we would find
from the above $g\simeq 10^{-30}$ which is unnaturally small. On
the other hand putting $\a_{YM}\sim O(10^{-2})$ we have ${R\over
l_I}\simeq 10^{-5}$. But when the compactification scale is
smaller than the string scale this  picture should not be used and
the physics should be analyzed in the T-dual situation. In this
case we have,
 \be
 G_N={1\over 8}{l_I^8g'^2\over
 R'^6},~~\a_{YM}=\shalf g'.
 \ee
Now ${R'\over l_I}\sim10^5$ and it would be very difficult to
understand how such large volume stabilizations could be
achieved.\footnote{It is perhaps also worth pointing out that
using a warp fact or in the metric with the transverse coordinates
allowed to take arbitrarily large values or even be non-compact as
is done in \cite{rs} does not solve this hierarchy problem.}.Also
the string coupling, although not unnaturally small,  is still
$O(10^{-1})$ and one would expect that perturbation theory is
valid and it is difficult to understand why the dilaton  should be
stabilized.

Note that the conclusion that $g'=\shalf \a_{YM}$ and is therefore
small in the second picture, is independent of the string scale
size or the compactification scale since these factors do not
enter into the above relation. In other words regardless of the
size of the string scale the IIB orientifold picture leads to a
weakly coupled string theory in which it is hard to understand how
the dilaton is stabilized. To avoid this situation we must argue
that the original picture is the correct one to use with $g\sim
O(1)$ and ${R\over l_I}\sim((2\a_{YM})^{-1/6}\sim 1.5$ so that it
is still reasonable to argue that the compactification scale is
close to the string scale and is thus presumably stabilized by
some SNP mechanism. It is not meaningful to go to the D3 brane
picture in this case since there $R'<l_I$ and the original picture
is the appropriate one.

The picture of string universality that we have presented would
then require that the region of moduli space where the real world
lies cannot be approached from this particular part of the
boundary of moduli space. This is not too surprising in any case.
The theory given in (\ref{tdualI}) is not of the same
status as the original 10D theories. It can be obtained  by
compactifying the type I theory on a torus (or orbifold) and then
T-dualizing. Now  according to our hypothesis if the original
type I theory  has a compactification that can approximate the
real world  in the central region of moduli space, just as its
S-dual heterotic theory can from the another side, this
compactification is likely to be a point in the moduli space of Ricci flat Kahler manifolds that is more complicated than a torus or
a orbifold  and this may not
necessarily be connected directly to the region of moduli space that yields
the type IIB orientifold. In other words  a
boundary region theory  of IIB orientifolds with D-branes
 may be  connected to the central region that we are interested in only via type I theories to which they connected through torus/orbifold compactifications and T-dualities. Otherwise we run into the problem of explaining why  the moduli are stabilized with small 10 D string coupling.

Now the scenario that we have considered here for a brane world
picture is a fairly conventional one and as we've argued above it
is hard to see how such a picture could work in that it requires moduli to be stabilized in the weak coupling region. An alternative
possibility is to have anti-branes as well as branes at orbifold
fixed points. Such an analysis has been carried out in 
\cite{Antoniadis:1999xk,AIQ}.
In this scenario SUSY is broken at the string scale which
therefore has to be taken to be  an intermediate one. The authors
have argued that there is possibly a mechanism for stabilizing the
moduli even at weak coupling. Even if this were the case these
models generically have a cosmological constant at the string
scale (there is no mechanism for generating a small number in the
theory and since everything is in principle calculable there is no
room for fine tuning either), and therefore would not solve the
practical cosmological constant problem.

It might be that the brane-world scenario requires a non-standard
compactification such as those which lead to gauged supergravity.
This is an issue which needs to be investigated further.

\section{Summary and Conclusions}

Practically all the calculations in string theory have been done
in the framework of one or other supersymmetric weakly coupled
theory  in the boundary region of moduli space. The problem that
needs to be addressed is what if anything these calculations tell
us about the real world which has:
\begin{itemize}
\item Four large space-time dimensions
\item No (or broken) supersymmetry.
\item Vanishing, or very small, cosmological constant.
\item A weak scale hierarchically smaller than the Planck scale
$M_P/M_Z\sim 10^{16}$, and perhaps comparable to the size of soft
supersymmetry breaking terms in the low energy theory.
\item Small gauge couplings.
\end{itemize}
A fundamental theory ought to be able to explain  these major
features of particle physics phenomenology. Unfortunately, at this
point we have no idea how string theory might explain them. As we
have reviewed at some length in the introduction, the source of
the  problem appears to be that the true vacuum of string theory
is in  the central region of moduli space where there are no
calculational techniques available.

Given this state of affairs, we have proposed a notion of string
universality which may provide a way of extracting information
about the central region from the information that we have in the
boundary region. The idea is based on the several assumptions. Of
our assumptions we believe that the first three are generally
accepted by string theorists:\\

({\em i})
 One underlying theory (M-theory?) exists, and has a connected
moduli space.\\

 ({\em ii})
The moduli space of this theory has boundary regions where the
theory approaches the various weakly coupled 10D string theories
and 11D supergravity.\\

({\em iii}) Non-perturbative effects in the theory  generate a
mechanism for stabilizing moduli, and thus fixing all parameters.
This provides a unique prediction of where the true vacuum
associated with the real world is situated in moduli space.\\

To these three assumptions we have added two more:\\

({\em iv}) The region in which moduli are stabilized is the
central region in which the string coupling is  of order unity and
the compactification scale is of order the string scale. This is
motivated by the duality relations between the different corners
of moduli space and  it is the region that is equally far in some sense from
these boundary regions\footnote{This assumption is similar to that
made  independently by G. Veneziano\cite{venezia}.}.\\

({\em v}) Some of the physics in this central region can be
extracted by approaching it from different calculable directions
and identifying the intersection of compatible predictions from
these extrapolations.\\

In section 1 and 2 of this paper we have discussed these
assumptions at some length. In section 3 we argued that the origin
of string non-perturbative effects are BPS brane instantons,
estimated their actions  and discussed their dualities. In Section
4 we highlighted the problems associated with currently popular
scenarios for moduli stabilization and SUSY breaking, and proposed
as an alternative that the two issues be decoupled. We propose
that moduli are all stabilized at or near the self-dual point by
string non-perturbative effects while SUSY breaking happens at a
much lower scale perhaps as a result of field theoretic
non-perturbative effects. In section 5 we have showed that
stabilization near the self dual point can be made consistent with
phenomenology.

We have concluded that moduli stabilization cannot occur when
either the inverse coupling or the volume are parametrically
large, and therefore at the very least, it is inconsistent to
consider, in ``race-track" models, only $T$ and $S$ moduli, and
parametrize SUSY breaking in terms of a complex vector in
$(F_T,F_S)$ plane. We believe that this is a very general
conclusion which shows how hard it is to make ``race-track" models
work.

We have concluded that moduli stabilization around the self-dual
point evades the problems of other scenarios. However, although
our ideas are not in conflict with phenomenology we have not yet
subjected them to a stringent test by constructing a concrete
model and verifying in a quantitative way that they are
consistent.

Finally, we would like to emphasize that we
have presented a new idea: String Universality, which then leads naturally
to many consequences, of which we have explored only some. Perhaps
the most important consequence of String Universality is that
the true vacuum of string theory is around the self-dual point with
string coupling of order unity and compact volume of the order of, but below
the string volume. We now need to confront this and find
new methods and models to describe the physics of string theory.

\section{Acknowledgments}
We wish to thank D. Eichler and G. Kane for discussions, and M.
Dine and Y. Shadmi for comments on the manuscript. We thank the
CERN TH group for hospitality during the time this project was
initiated. This work is partially supported by the Department of
Energy contract No. DE-FG02-91-ER-40672.


\end{document}